\documentclass{epl}
\usepackage{psfig}
\begin{document}

\title{Multiplet effects in the electronic structure of $\delta$-Pu, Am
      and their compounds}
\shorttitle{Multiplet effects in $\delta$-Pu, Am}
\author{Alexander Shick\inst{1}, Jind\u{r}ich Koloren\v{c}\inst{1,2}, Ladislav
Havela\inst{3}, V\'aclav Drchal\inst{1}, and Thomas Gouder\inst{4}}
\institute{
  \inst{1} Institute of Physics, ASCR, Na Slovance 2, CZ-18221 Prague, Czech Republic\\
  \inst{2} Department of Physics, North Carolina State University, Raleigh, NC 27695\\
  \inst{2} Department of Condensed Matter Physics, Faculty of Mathematics and Physics, Charles University,
  Ke Karlovu 5, CZ-12116 Prague, Czech Republic\\
  \inst{3} European Commission, Joint Research Centre, Institute for
  Transuranium Elements, Postfach 2340, 76 125 Karlsruhe, Germany
}

\date{\today}
\maketitle

\vspace*{-0.75cm}
\begin{abstract}
We propose a straightforward and efficient procedure to perform
dynamical mean-field (DMFT) calculations on the top of the static
mean-field LDA+U approximation. Starting from self-consistent LDA+U
ground state we included multiplet transitions using the Hubbard-I
approximation, which yields a very good agreement with experimental
photoelectron spectra of $\delta$-Pu, Am, and their selected
compounds.
\end{abstract}

\vspace*{-1.25cm}

\section{Introduction}
Electronic, magnetic and superconducting properties of actinide
elements recently attracted significant interest and attention in
the condensed matter physics.
Most intriguing are the phenomena at the  localization
threshold of the $5f$ series, which is crossed between Pu and Am,
where the electron-electron correlations play a prominent role
\cite{Kotliar,grivenau}.

During last few years, electronic structure calculations of Pu and Am
based on the conventional band theoretical
methods (the local density or generalized
gradient expantion approximations LDA/GGA  to the density
functional theory) could not explain essental experimental data. While, the LDA/GGA band
structure calculations predict a local magnetic moment (ordered or
disordered) to form on the Pu \cite{soderlind} and Am
\cite{soderlind2} atoms, none of them were seen in the experiment
\cite{lashley}. Also, the same papers attempted to evaluate the
photoemission spectra (PES) and electronic specific heat in Pu and
Am making use of single-particle LDA/GGA densities of states (DOS),
incorrectly assuming weak electron correlation character of 5$f$
systems at the borderline between the localized, nonbonding,
behaviour and the bonding situation of electronic bands.

It was shown recently that the {\em around-mean-field} (AMF)-LSDA+U
correlated band theory gives non-magnetic ground state for Pu
\cite{shick05} and Am \cite{shick06}. Also, the equilibrium volumes
and bulk moduli for $\delta$-Pu and {\em fcc-}Am are calculated in a
good agreement with experiment.  However, there is a clear
disagreement between the AMF-LDA+U calculated DOS and PES,
questioning the validity of this approximation. The DMFT
calculations \cite{EPLP} using the (AMF)-LSDA+U form of interacting
Hamiltonian demonstrated that experimental PES and high $\gamma$
coefficient of the electronic specific heat in $\delta$-Pu and its
selected compounds \cite{PLK05,PKL06} originate from the
excitations, and not from the ground state DOS alone.
While accounting for dynamical fluctuations, the DMFT of Ref.
\cite{EPLP} does not take into account atomic-like excitations which
can play an important role in Pu and Am.

In this Letter, we develop a DMFT based computational scheme based
on multi-orbital Hubbard-I approximation (HIA)
\cite{LK99,Dai03,Dai05} including the spin-orbit coupling (SOC)
which explicitly accounts for the atomic-like multiplet transition
excitations in Pu, Am and their compounds. Starting from the
non-magnetic ground state calculated with the static mean-field
AMF-LSDA+U approximation, we obtain excitation spectra of Pu and Am
in surprisingly good agreement with PES, in support of the
atomic-like origin of the electronic excitations in these materials.

\section{Methodology}
We start with the multi-band Hubbard Hamiltonian \cite{LK99}
$H = H^0 + H^{\rm int} $, where
\begin{eqnarray}
\label{eq:1ph} H^0 = \sum_{i,j} \sum_{\alpha, \beta} H^0_{i \alpha,
j \beta}
                 c^{\dagger}_{i \alpha} c_{j \beta}
    = \sum_{\bf k} \sum_{\alpha, \beta} H^0_{\alpha, \beta} ({\bf k})
        c^{\dagger}_{\alpha}({\bf k}) c_{\beta}({\bf k}) \, ,
\end{eqnarray}
is the one-particle Hamiltonian found from ab initio electronic
structure calculations for a periodic crystal. The indices $i,j$
label the lattice sites, $\alpha = (\ell m \sigma)$ denote the
spinorbitals, and ${\bf k}$ is the k-vector from the first Brillouin
zone. It is assumed that the electron-electron correlations between
s, p, and d electrons are well described within the density
functional theory, while the correlations between the f electrons
have to be considered separately by introducing the interaction
Hamiltonian
\begin{eqnarray}
\label{eq:hint} H^{\rm int} = \frac{1}{2} \sum_i
\sum_{m_1,m_2,m_3,m_4}
                          \sum_{\sigma,\sigma'}
  \langle m_1,m_2|V_{i}^{ee}|m_3,m_4 \rangle
  c_{i m_1 \sigma}^{\dagger} c_{i m_2 \sigma'}^{\dagger}
  c_{i m_4 \sigma'} c_{i m_3 \sigma}  \, .
\end{eqnarray}
The $V^{ee}$ is an effective on-site Coulomb interaction \cite{LK99}
expressed in terms of the Slater integrals $F_k$ and the spherical
harmonic ${|lm \rangle}$. The corresponding one-particle Green function
\begin{eqnarray}
\label{eq:1gf} G({\bf k},z) = \Big( z + \mu - H^0 ({\bf k}) - \Sigma({\bf k},z) \Big)^{-1}
\end{eqnarray}
is expressed via $H^0$ and the one-particle selfenergy $\Sigma({\bf k},z)$
which contains the electron-electron correlations,
where $z$ is a (complex) energy with respect to the chemical
potential $\mu$. The interactions
(\ref{eq:hint}) act only in the subspace of f-states. Consequently,
the selfenergy $\Sigma({\bf k},z)$ is nonzero only in the subspace of the
f-states.

%

The simplest mean-field approximation (often called L(S)DA+U)
\cite{LAZ95} neglects the ${\bf k}$- and energy-dependence of
$\Sigma$ replacing it by the on-site potential $V_{+U}$. For a given
set of spin-orbitals $|m \sigma \rangle$ the potential reads:
\begin{eqnarray}
[V_{+U}]^{\sigma}_{mm'} = \sum_{p,q,\sigma'} \Big( \langle
m,p|V^{ee}|m',q \rangle - \langle m,p|V^{ee}|q,m' \rangle
 \delta_{\sigma,\sigma '} \Big) n^{\sigma'}_{p,q} \, ,
\label{eq:4}
\end{eqnarray}
where $n^{\sigma}_{mm'}$ is the local orbital occupation matrix of
the orbitals $|m \sigma \rangle$. It was shown in Ref.
\cite{SLP99} that the Kohn-Sham equation with the potential
Eq.(\ref{eq:4}) can be obtained by making use of variational
minimization of the LDA+U total energy functional (i.e. of the
expectation value of the multiband Hubbard Hamiltonian) in a way
similar to the conventional density functional theory \cite{PZ81}.

In what follows we use the local approximation for the selfenergy,
i.e., we assume that it is site-diagonal and therefore independent
of ${\bf k}$. Then we can employ the ``impurity" method of Ref.
\cite{LK99}. We first obtain a local Green function integrating $G({\bf k},z)$,
Eq.(\ref{eq:1gf}), over the Brillouin zone
\begin{eqnarray}
G(z) = \frac{1}{V_{BZ}} \int_{BZ}d{\bf k}
\Big( z + \mu - H^0 ({\bf k}) - \Sigma(z) \Big)^{-1}
\label{eq:gff}
\end{eqnarray}
and define  a ``bath" Green function (the
so-called Weiss field) ${\cal  G}_{0}(z)$
\begin{eqnarray}
{\cal  G}_{0}(z) = \Big( G^{-1}(z) + \Sigma(z) \Big)^{-1} \, .
\label{eq:6}
\end{eqnarray}
All ${\cal G_0}$, $G$, and $\Sigma$ in Eqs. (\ref{eq:gff}, \ref{eq:6})
are matrices in the subspace of the $|m \sigma \rangle$ f-orbitals.

The DMFT self-consistency condition is now formulated by equating $G
(z)$, Eq.(\ref{eq:gff}), to the Green function $\tilde{G}(z)$ of a
single-impurity Anderson model (SIAM) \cite{PWN97}
\begin{eqnarray}
\label{eq:8} \tilde{G}(z) = \Big( z + \mu - \epsilon_0 -
\tilde{\Delta}(z) -
  \tilde{\Sigma}(z) \Big)^{-1}
\end{eqnarray}
where $\tilde{\Delta}(z)$ is the effective hybridization function
and $\tilde{\Sigma}(z)$ is the SIAM self-energy. We write the
Eq.(\ref{eq:8}) in the form of the Eq.(\ref{eq:6})
\begin{eqnarray}
\Big( z + \mu - \epsilon_0 - \tilde{\Delta}(z) \Big)^{-1}
  = \tilde{{\cal G}}_0(z)
  = \Big( \tilde{G}^{-1}(z) + \tilde{\Sigma}(z) \Big)^{-1}
\label{eq:9}
\end{eqnarray}
from which it follows that $\tilde{\cal G}_0(z)$ has the meaning of
the SIAM ``bath" Green function. The iterative procedure to solve
the periodic lattice problem in the DMFT approximation is now
formulated in a usual way \cite{LK99}: starting with single particle
Hamiltonian $H^0({\bf k})$ and a guess for local $\Sigma$, the local
Green function is calculated using the Eq.(\ref{eq:gff}) and the
``bath" Green function is calculated from Eq.(\ref{eq:6}); the SIAM
is solved for this ``bath" and a new local $\Sigma$ is calculated
from Eq.(\ref{eq:8}), which is inserted back into Eq.(\ref{eq:gff}).

The LDA+U procedure can be viewed in the same way. There is no need
to apply the full DMFT iterative procedure described above and to
solve the SIAM with the ``bath" from Eq.(\ref{eq:6}). The
self-energy $\Sigma(z)$ is now approximated by a static potential
$V_{+U}$ from Eq.(\ref{eq:4}), and the well-known relation between
the Green function, Eq.(\ref{eq:6}), and the local orbital
occupation matrix $n^{\sigma}_{mm'} \, = \, -\pi^{-1}{\rm Im}
\int^{\mu} {\rm d} E \,  G(E)^{\sigma}_{mm'}$ is used. In addition,
the charge- and spin-densities needed to construct the
single-particle Hamiltonian $H_0({\bf k})$ in Eq.(\ref{eq:gff}) are
calculated self-consistently. We emphasize that LDA+U approximation
is generically connected with the LDA+DMFT procedure.

\section{Hubbard-I approximation}
Here we attempt to build a computational scheme based on
self-consistent static mean-field LDA+U ground state that will allow
us to access the correlated electron excitations. We specifically
choose the {\em around-mean-field} version of LSDA+U (AMF-LDA+U)
which was shown to describe correctly the non-magnetic ground state
properties of $\delta$-Pu \cite{shick05}, fcc-Am, Pu-Am alloys
\cite{shick06}, and selected Pu-compounds \cite{PLK05}.
We extend our previous works \cite{shick05,shick06} towards the DMFT
to account for the multiplet transitions which are necessary for a
correct description of PE excitation spectra. We use the
multiorbital HIA which is suitable for incorporating the multiplet
transitions in the electronic structure, as it is explicitly based
on the exact diagonalization of an isolated atomic-like f-shell.

Further, we restrict our formulation to the paramagnetic phase, and
we closely follow the procedure described in \cite{LK99}. We
construct the atomic Hamiltonian including the spin-orbit coupling
(SOC):
\begin{eqnarray}
H^{\rm at} = \sum_{m_1,m_2}^{\sigma, \sigma'}
 \xi ({\bf l} \cdot {\bf s})_{m_1 m_2}^{\sigma \; \; \sigma'}
 c_{m_1 \sigma}^{\dagger}c_{m_2 \sigma'}
+ \frac{1}{2} \sum_{m_{1}...m_{4}}^{\sigma, \sigma'} \langle
m_1 m_2|V^{ee}|m_3 m_4 \rangle c_{m_1 \sigma}^{\dagger} c_{m_2
\sigma'}^{\dagger} c_{m_4 \sigma'} c_{m_3 \sigma} \, , \label{eq:10}
\end{eqnarray}
and perform exact diagonalization of $H^{\rm at} |\nu \rangle =
E_\nu |\nu\rangle $ to obtain all possible eigenvalues  $E_\nu$ and
eigenvectors $|\nu \rangle$.

The HIA ``chemical potential"  $\mu_{H}$ is then calculated as
\begin{equation}
\langle n \rangle = \frac{1}{Z} {\rm Tr} \Big[ N \exp(-\beta [H^{\rm
at}
       - \mu_H \hat{N}]) \Big]
\label{eq:11}
\end{equation}
for a given number of particles $\langle n \rangle$. Here, $\beta$
is the inverse temperature and $Z$ is the partition function.

Finally, the atomic Green function is calculated as follows:
\begin{eqnarray}
[G^{\rm at}(z)]_{m_1 m_2}^{\sigma \; \; \sigma'}(z) &=& \frac{1}{Z}
\, \sum_{\nu,\mu} \frac{\langle \mu|c_{m_1 \sigma}|\nu \rangle
\langle \nu|c_{m_2 \sigma'}^{\dagger}|\mu \rangle} {z + (E_\mu -
\mu_H N_{\mu}) - (E_\nu - \mu_H N_{\nu})}
\nonumber \\
&& \times [\exp(-\beta (E_\nu - \mu_H N_\nu))
         + \exp(-\beta (E_\mu - \mu_H N_\mu))]
\label{eq:12}
\end{eqnarray}
and the atomic self-energy is evaluated as:
\begin{eqnarray}
\label{eq:13} [\Sigma_{H}(z)]^{\sigma \; \; \sigma'}_{mm'} =
 z \delta_{m_1 m_2} \delta_{\sigma \sigma'} - \xi ({\bf l} \cdot
{\bf s})_{m_1 m_2}^{\sigma \; \; \sigma'}  - \Big[ \Big(G^{\rm
at}(z)\Big)^{-1}\Big]_{m_1 m_2}^{\sigma \; \; \sigma'} \, .
\end{eqnarray}

Assuming that the self-consistent LDA+U calculations are performed,
the LDA+U Green function is evaluated as
\begin{eqnarray}
G_{+U}(z) = \frac{1}{V_{\rm BZ}} \int_{\rm BZ}d \, {\bf k}
\Big(z+\mu - H_0({\bf k}) - V_{+U} \Big)^{-1} \, . \label{eq:14}
\end{eqnarray}
In Eq.(\ref{eq:14}) we took into account the presence of SOC for
both  $H_0({\bf{k}})$ and $\hat{V}_{+U}$
as described in Ref. \cite{shick05}. We used LDA+U
eigenvalues and eigenfunctions calculated in the full-potential LAPW
basis \cite{SLP99,shick05} to construct the on-site spin-orbital Green
function matrix, Eq.(\ref{eq:14}).

We evaluate the static ``bath" ${\cal G}_{0}(z)$ from
Eq.(\ref{eq:6}) and find the hybridization function $\Delta(z)$
together with $( \epsilon_0 - \mu)$ which determines the energy  of
``impurity" level with respect to the solid potential:
\begin{eqnarray}
\Big( z + \mu - \epsilon_{+U} - \xi ({\bf l} \cdot {\bf s}) -
 \Delta(z) \Big)^{-1}  =  {\cal G}^0_{+U}(z)
= \Big( G_{+U}^{-1}(z) + V_{+U}(z) \Big)^{-1} \, . \label{eq:15}
\end{eqnarray}
Here we added the SOC explicitly and assumed the paramagnetic case.

We point on a difference in a physical meaning of $\epsilon_0$ in
the original SIAM and in the auxiliary SIAM used in LDA+DMFT
(LDA+U): in the former, it labels the position of the
non-interacting f(d)-level so that the chemical potential is
determined self-consistently for a given $\Delta(z)$; in the latter,
the chemical potential $\mu$ is determined by a periodical crystal -
basically by the Green function of Eq.(\ref{eq:gff}) - so that the
auxiliary $\epsilon_0$ accommodates all the electrostatic shifts
between correlated electrons and the potential of the solid.

Now we can formulate a simple approximate procedure to solve the
DMFT Eqs.(\ref{eq:gff},\ref{eq:6},\ref{eq:8}) with Hubbard-I
self-energy, Eq.(\ref{eq:13}). We assume that the self-consistent
static mean-field LDA+U already gives a correct number of particles
and hybridization. Using the LDA+U Green function, Eq.(\ref{eq:10}),
and the potential, Eq.(\ref{eq:4}), we evaluate the Weiss field
${\cal G}_{+U}^0(z)$. Then we insert the HIA self-energy, calculated
for the same number of correlated electrons as given by LDA+U, into
this ``bath", and calculate the new Green function
\begin{eqnarray}
G(z) =  \Big( [{\cal G}_{0}^{+U}(z)]^{-1} + (\epsilon_{+U} -
\epsilon_{H}) - \Sigma_{H}(z) \Big)^{-1} \, , \label{eq:16}
\end{eqnarray}
where $(\epsilon_{+U} - \epsilon_{H})$ is chosen so as to ensure
that $n \; = \; \pi^{-1} {\rm Im} \int^{\mu} d E {\rm Tr}[ G(E)]$ is
equal to a given number of correlated electrons \cite{shift}.

\section{Valence band photoemission spectra}
Starting from self-consistent AMF-LSDA+U ground state solutions for
${\delta}$-Pu \cite{shick05}, and \textit{fcc}-Am \cite{shick06},
and making use of the corresponding eigenvalues and eigenfunctions
we evaluate the $G_{+U}(z)$ Green function given by Eq.(\ref{eq:14})
and the ${\cal G}_{0}^{+U}(z)$  ``bath" Green Function
(Eq.(\ref{eq:15})) \cite{details}. In HIA calculations
Eq.(\ref{eq:10} - \ref{eq:13}) we used the commonly accepted values
of SOC constants $\xi$ = 0.3 eV (Pu) and 0.34 eV (Am). To find
$\mu_H$ Eq.(\ref{eq:11}), we choose the AMF-LDA+U values of $\langle
n \rangle$, namely, $n_f$=6.0 for Am and $n_f$=5.4 for Pu
\cite{details2}.  The self-energy Eq.(\ref{eq:13}) was calculated
along the real axis for $z=E-E_{\rm F} + i \delta$, where $\delta$ =
63 meV \cite{Svane},
 and $\beta$ was varied from 100 to 1000
eV$^{-1}$. We found no sizable effect due to the variation of
$\beta$ for a given $\mu_H$ in the resulting spectral density
Eq.(\ref{eq:16}).

In Fig.1a we show the f-projected DOS (fDOS) from AMF-LDA+U together
with spectral density calculated from Eq.(\ref{eq:16}). For {\em
fcc}-Am, the well localised fDOS peak at  -4 eV transforms to the
multiplet of excited state transitions $f^6 \rightarrow  f^5$ below
the Fermi energy, and fDOS-manifold around +2 eV to $f^6 \rightarrow
f^7$  multiplet transitions. Although there is no doubt that the 5f
multiplets must dominate the experimental valence-band spectra of
Am-based systems, individual lines are not resolved (except for
partly resolved features in  the spectrum of Am metal), and the
position of the 5f intensity in the energy spectrum is the main
indicator of the agreement with calculations. In this sense, the
calculated Am spectral density is in a good agreement with PES
\cite{gouder05}. It also agrees with similar calculations
\cite{Svane} \cite{note}. Furthermore, we performed the
calculations of AmN and AmSb using the same $\{U,J\}$ set of values
as for elemental Am. The AMF-LDA+U yields AmN as an indirect gap
semiconductor and AmSb as a semi-metal. The HIA spectral densities
for Am $f$ manifolds in AmN and AmSb are shown in Fig.1b,c and are
in good agreement with PES experiments \cite{gouder05}.

Experimental valence-band spectra of $\delta$-Pu and several other
Pu systems exhibit three narrow features within 1 eV below $E_F$,
the most distinct one very close to $E_F$ being accompanied by a
weaker feature at 0.5 eV and another one at 0.8-0.9 eV. Their
general occurrence and invariability of characteristic energies
practically excludes any relation to individual features in density
of electronic states. Instead, a relation to final state multiplets
has been suggested \cite{gouder00,havela02} among other possible
explanations . Similar to Am, the link to atomic multiplet is
corroborated by the present calculations also for Pu. For
$\delta$-Pu shown in Fig.2a, the AMF-LDA+U fDOS manifold at around
-1 eV transforms into a set of multiplet transitions with high value
of spectral density at $E_F$. Similarly to the Pu metal,  the PES
exhibits the three most intense peaks for a broad class of Pu
compounds. The calculations performed for PuTe (see Fig. 2b) indeed
demonstrate the three-peak pattern similar to Pu, in agreement
with experiment \cite{durakiewicz}. Also we point out a good
agreement between our calculations and recent DMFT calculations of
Ref. \cite{EPLP}, as well as those of Svane for PuSe \cite{Svane}.
One should note that the high spectral intensity at the Fermi level
is explaining enhanced values of the $\gamma$-coefficient of
electronic specific heat, observed in $\delta$-Pu and other Pu
systems.

As a conclusion, we have shown that the three narrow features
observed in the valence band spectra of Pu and majority of Pu
compounds can be identified with the most intense atomic excitations
(multiplets), calculated using the LDA+U and the multi-orbital HIA
calculations. The calculations explain that the atomic excitations
can be observed even if the 5f states are not fully localized as in
$\delta$-Pu, and the atomic character fixes the characteristic
energies (not intensities) such that similar features are found in
spectra of diverse Pu systems. A reasonable agreement with
experiment is found also for Am and its compounds calculated on the
same footing.

The research was carried out as a part of research programs
AVOZ10100520 of the Academy of Sciences, MSM 0021620834 of the
Ministry of Education of Czech Republic. Financial support was
provided by the Grant Agency of the Academy of Sciences (Project
A100100530), the Grant Agency of Czech Republic (Projects
202/04/1005 and 202/04/1103), and by the action COST P16 (Project
OC144, financed by the Czech Ministry of Education). We gratefully
acknowledge valuable discussions with V. Jani\v{s}, P. Oppeneer, and
A. Lichtenstein.

\vspace*{-1cm}
\begin{figure}[t]
\twoimages[scale=1.10]{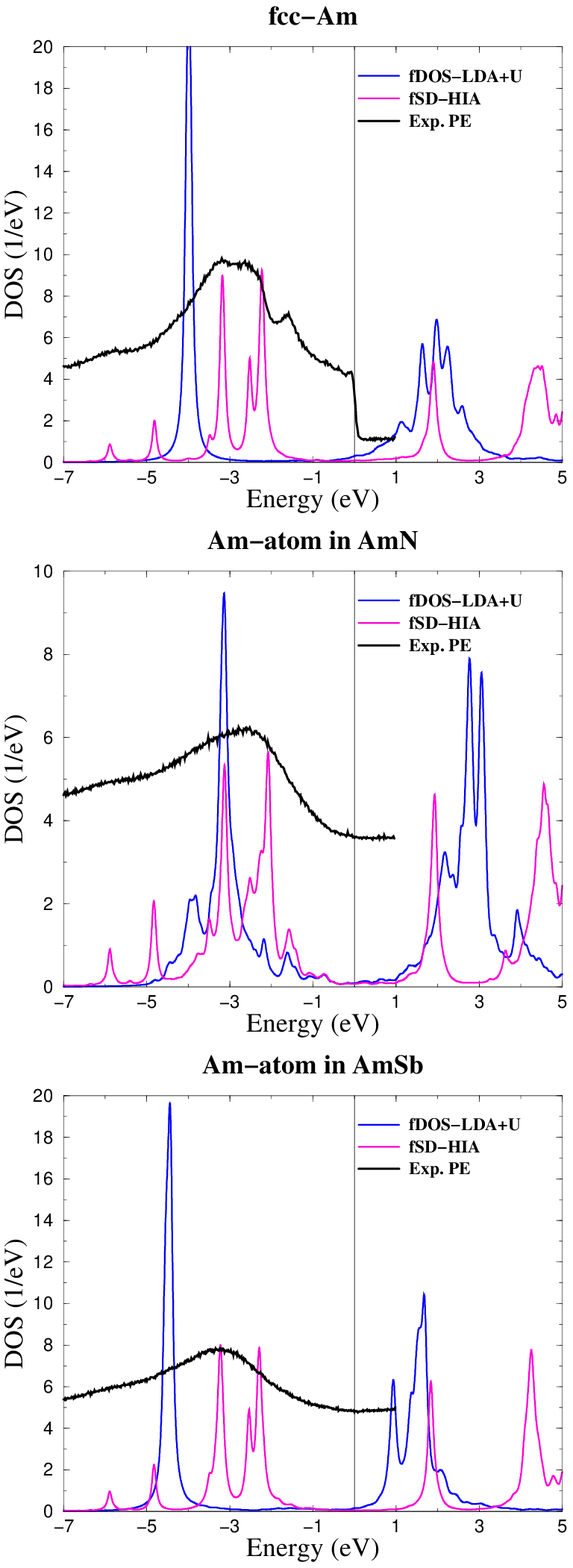}{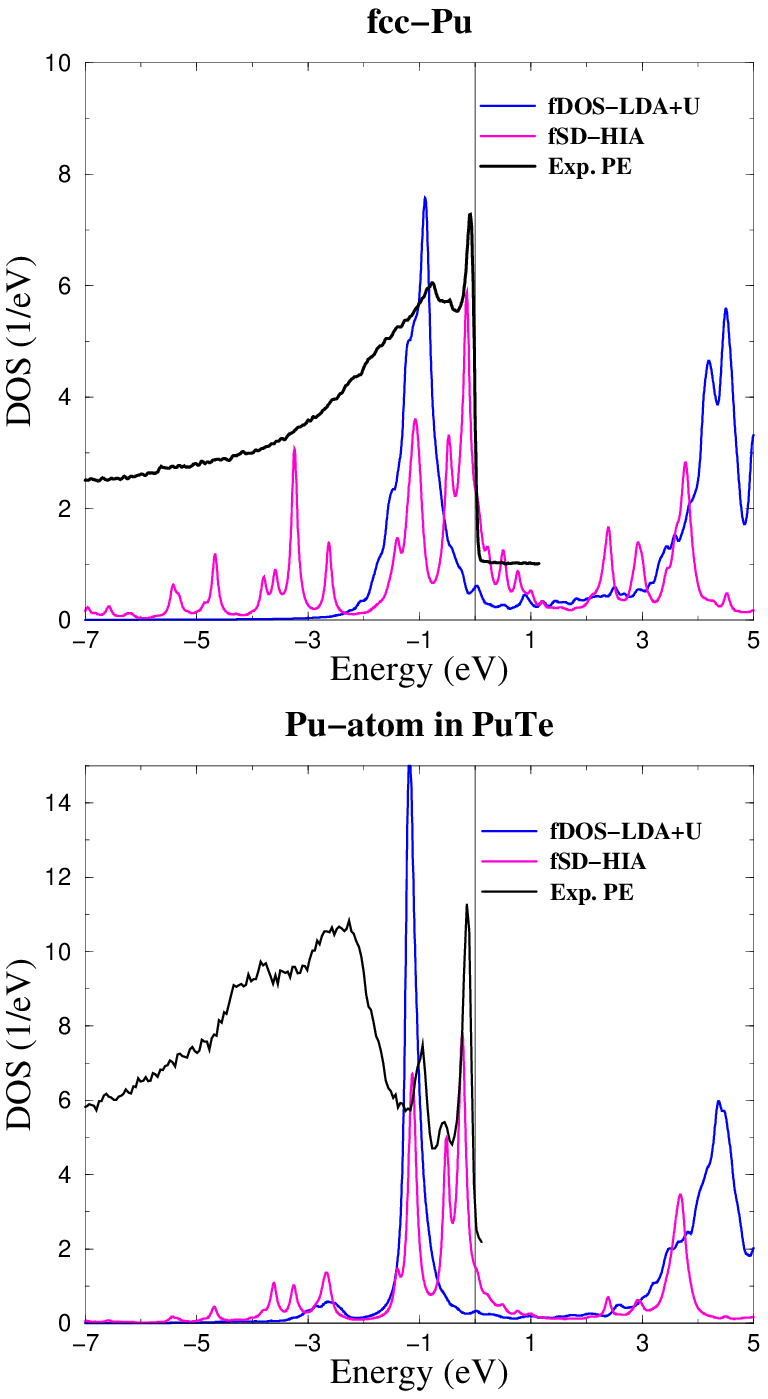}

\vspace*{-0.5cm} \caption{Color-on-line. The fDOS (AMF-LDA+U) and
Spectral Density (HIA) for \textit{fcc}-Am (top-a), AmN (middle-b),
AmSb(bottom-c) calculated along $z=E-E_F + i \delta$ where $\delta$
= 63 meV is used. Experimental PES (arb. units) from Ref.
\cite{gouder05} are shown \cite{PES}.}

\vspace*{-0.25cm} \caption{Color-on-line. The fDOS (AMF-LDA+U) and
Spectral Density (HIA) for  $\delta$-Pu (top-a) and PuTe (bottom-b)
along $z=E-E_F + i \delta \; (\delta)$= 63 eV. Experimental PES
(arb. units) from Ref. \cite{havela02} ($\delta$-Pu) and from Ref.
\cite{durakiewicz} (PuTe) are given \cite{PES}.}
\end{figure}


\begin{thebibliography}{99}
\bibitem{details2}
We found that the Eq.(\ref{eq:11}) gives practically the same values
for $\mu_H$ for inverse temperature $\beta = 1/ (k_B T)$ in the
range of $30-50$ eV$^{-1}$ as the procedure of Ref. \cite{LK99}
based on the summation over Matsubara frequencies (see
Eq.(\ref{eq:12}) of Ref. \cite{LK99}).
\bibitem{PES} The experimental spectra were shifted upwards for
clarity.
\end{thebibliography}
\end{document}